\documentclass[twocolumn]{aastex6}
\pdfoutput=1 
\usepackage{amsmath,amstext}
\usepackage[T1]{fontenc}
\usepackage{apjfonts} 
\usepackage[figure,figure*]{hypcap}
\usepackage{affils}
\usepackage{hyperref}
\usepackage{microtype}
\usepackage{caption}
\usepackage{subcaption}
\usepackage{xspace}

\def\M{M$_\odot$}
\def\Z{Z$_\odot$}
\def\kms{km\,s$^{-1}$}

\def\ra#1#2#3{#1$^{\rm h}$#2$^{\rm m}$#3$^{\rm s}$}
\def\dec#1#2#3{#1$^\circ$#2$'$#3$''$}

\def\mosfit{\texttt{MOSFiT}\xspace}

\shorttitle{SN\,2017egm: a SLSN in NGC\,3191}
\shortauthors{Nicholl et al.}

\begin{document}

\title{The superluminous supernova SN\,2017egm in the nearby galaxy NGC\,3191: a metal-rich environment can support a typical SLSN evolution}

\DeclareAffil{cfa}{Harvard-Smithsonian Center for Astrophysics, 60 Garden Street, Cambridge, Massachusetts 02138, USA; \href{mailto:matt.nicholl@cfa.harvard.edu}{matt.nicholl@cfa.harvard.edu}}
\DeclareAffil{northwestern}{Center for Interdisciplinary Exploration and Research in Astrophysics (CIERA) and Department of Physics and Astronomy, Northwestern University, Evanston, IL 60208}
\DeclareAffil{ohio}{Astrophysical Institute, Department of Physics and Astronomy, 251B Clippinger Lab, Ohio University, Athens, OH 45701, USA}

\affilauthorlist{Matt Nicholl\affils{cfa},
Edo Berger\affils{cfa}, Raffaella Margutti\affils{northwestern}, Peter K.~Blanchard\affils{cfa}, James Guillochon\affils{cfa}, Joel Leja\affils{cfa} and Ryan Chornock\affils{ohio}
}

\begin{abstract}
At redshift $z=0.03$, the recently-discovered SN\,2017egm is the nearest Type I superluminous supernova (SLSN) to date, and first near the center of a massive spiral galaxy (NGC\,3191). Using SDSS spectra of NGC\,3191, we find a metallicity $\sim2$\,\Z\ at the nucleus and $\sim1.3$\,\Z\ for a star forming region at a radial offset similar to SN\,2017egm. Archival radio-to-UV photometry reveals a star-formation rate $\sim15$\,\M\,yr$^{-1}$ (with $\sim70$\% dust-obscured), which can account for a \textit{Swift} X-ray detection, and stellar mass $\sim10^{10.7}$\,\M. We model the early UV-optical light curves with a magnetar central-engine model, using the Bayesian light curve fitting tool \texttt{MOSFiT}. The fits indicate ejecta mass $2-4$\,\M, spin period $4-6$\,ms, magnetic field $(0.7-1.7)\times10^{14}$G, and kinetic energy $1-2\times10^{51}$\,erg. These parameters are consistent with the overall distributions for SLSNe, modeled by \citet{nic2017c}, although the derived mass and spin are towards the low end, possibly indicating enhanced loss of mass and angular momentum before explosion. This has two implications: (i) SLSNe can occur at solar metallicity, although with a low fraction $\sim10$\%; and (ii) metallicity has at most a modest effect on their properties. Both conclusions are in line with results for long gamma-ray bursts. Assuming a monotonic rise gives an explosion date MJD\,$57889\pm1$. However, a short-lived excess in the data relative to the best-fitting models may indicate an early-time `bump'. If confirmed, SN\,2017egm would be the first SLSN with a spectrum during the bump-phase; this shows the same \ion{O}{2} lines seen at maximum light, which may be an important clue for explaining these bumps.
\end{abstract}

\keywords{supernovae:general --- supernovae:individual(SN\,2017egm)}

\section{Introduction}

Modern time-domain surveys have become adept at uncovering new and rare types of transients. One of the most important findings has been the discovery of hydrogen-poor stellar explosions that reach peak absolute magnitudes $M\lesssim -21$ mag \citep{qui2011,chom2011,gal2012} -- the so-called Type I superluminous supernovae (here referred to simply as superluminous supernovae or SLSNe).  To date, these have been found almost exclusively in metal-poor dwarf galaxies at redshifts $z\approx 0.1-2$ \citep{lun2014,lel2015,per2016,chen2016,ang2016,schu2016}. SLSNe span a wide range of luminosities and timescales, but are uniquely defined by a blue spectrum at early time (with blackbody temperature $\gtrsim 15000$\,K), which later cools to resemble more typical Type Ic SNe \citep{pas2010,ins2013,nic2015b}.

While the power source of SLSNe was initially a great mystery, a wide range of recent studies point to a `central engine', namely the rotational power of a central compact remnant.  Given that the luminosity of SLSNe remains high for weeks to months, spin-down power from a millisecond magnetar \citep{kas2010,woo2010,met2015} is more plausible than black hole accretion \citep{dex2013}. \citet{nic2017c} recently presented magnetar model fits to the multicolor light curves of 38 SLSNe to uniformly determine the parameter space occupied by these explosions and their magnetar engines, using our new public Bayesian code \mosfit. Key findings of this work include a continuous, and relatively narrow distribution of the SN and magnetar parameters, and a lack of dependence on metallicity.

Here, we use the same framework to study the early evolution of SN\,2017egm in the first 50 days after explosion. SN\,2017egm is the nearest SLSN to date ($z=0.0307$, $d_L=135$ Mpc) and the first to be robustly associated with a nearby massive spiral galaxy, NGC\,3191 \citep{dong2017}. Our goals are threefold: (i) determine the time of explosion and the SN and engine properties; (ii) determine the properties of the host, NGC\,3191; and (iii) explore how SN\,2017egm fits into the broader SLSN class in light of its unusual host environment.  The former will help to focus searches for any pre-peak `bumps' as seen in several previous SLSNe \citep{lel2012,nic2015a,nic2016a,smi2016}, and to inform on-going follow-up strategies. The latter two points will shed light on the nature of the engine and SN and their relation (if any) to the environment's metallicity, as well as on the comparison of SLSNe to long gamma-ray bursts (LGRBs).

\section{SN\,2017egm: a low-redshift SLSN}

\begin{figure}
\includegraphics[width=\columnwidth]{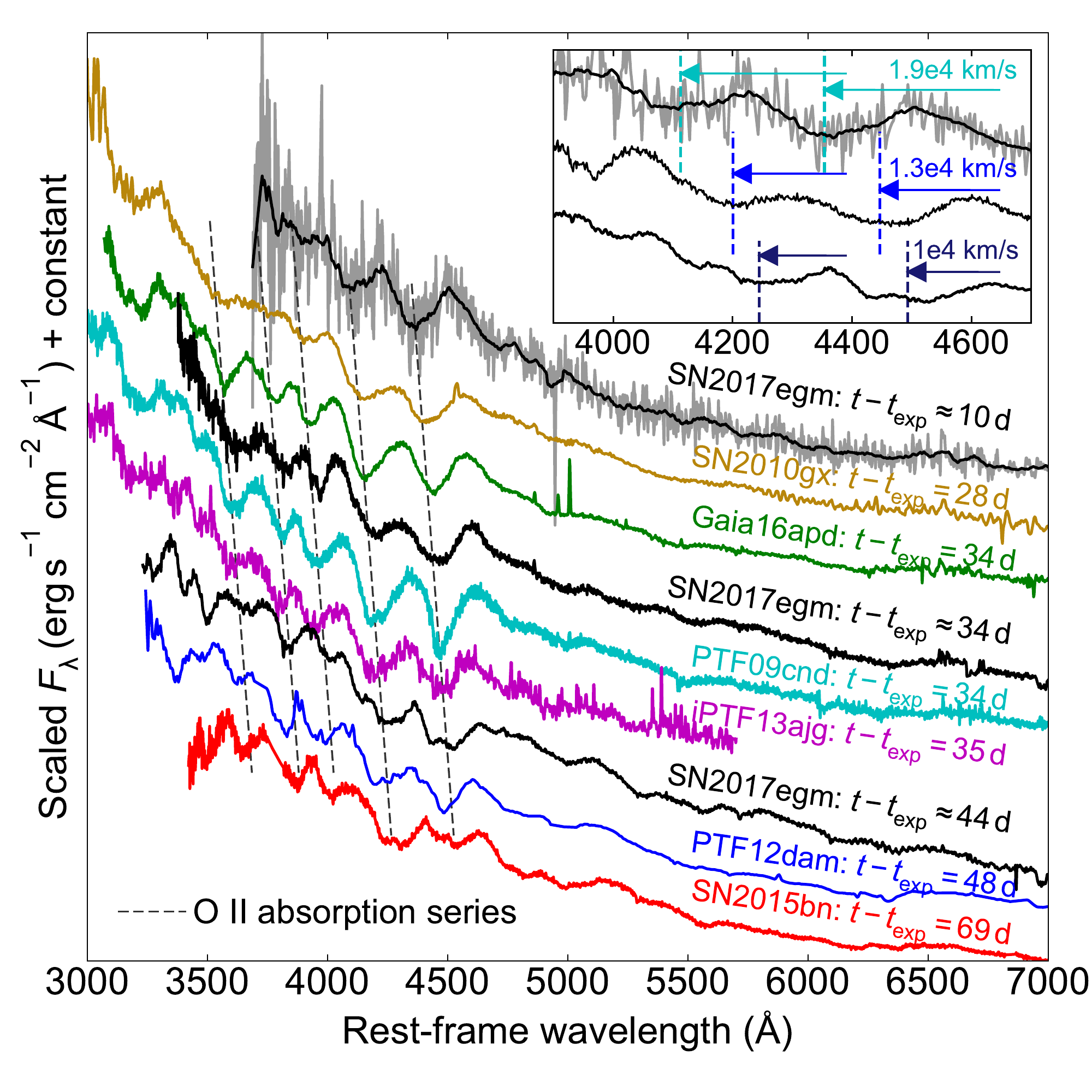}
\caption{Early-phase spectra of SN\,2017egm (black) compared to pre-maximum spectra of several well-studied SLSNe from the literature. The spectra are arranged in order of phase from explosion. Dashed lines indicate the positions of the signature \ion{O}{2} lines, showing the systematic decrease in absorption line velocity at early times. Data for the literature objects are from \citet{pas2010,nic2017,vre2014,qui2011,nic2013,nic2016b}}
\label{fig:spec}
\end{figure}

\subsection{Discovery}

SN\,2017egm was discovered by the \textit{Gaia} Satellite on 2017 May 23 UT and given the internal designation Gaia17biu. It was not detected in previous observations on 2017 Apr 20. It was reported to the Transient Name Server\footnote{https://wis-tns.weizmann.ac.il/} by the Gaia Photometric Science Alerts team \citep{del2017}. The SN is coincident with the star-forming galaxy NGC\,3191, with a radial offset of $5.2''$ from the nucleus (see \S\ref{sec:host}).

An optical spectrum was obtained by \citet{xia2017} on 2017 May 26.6 using the 2.16\,m telescope at Xinglong Station of the National Astronomical Observatories of China. The spectrum was blue with only weak features, and the authors found a best match to a young Type II SN using the SuperNova Identification Code \citep{blo2007}; however they also noted that the absolute magnitude, $\approx -19$\,mag, was unusually high for this spectral type.

\citet{dong2017} re-observed SN\,2017egm at higher signal-to-noise on 2017 May 30 with the Nordic Optical Telescope as part of the NOT Unbiased Transient Survey.  They found broad absorption features matching the \ion{O}{2} series common in SLSNe.  Photometry subsequently obtained with the \textit{Swift} UV Optical Telescope (UVOT) showed an absolute magnitude of $M_V\approx -20.6$ and a colour temperature of $\gtrsim 15,000$\,K, confirming that SN\,2017egm is a SLSN at a distance roughly half that of the previous nearest event \citep{per2016}. \citet{dong2017} also noted that the massive spiral host galaxy is very unusual for SLSNe.

\subsection{Early Spectra}

We observed SN\,2017egm with the FAST spectrograph \citep{tok1997,fab1998} on the 60-inch telescope at Fred Lawrence Whipple Observatory (FLWO) on 2017 June 18 and with MMT using the Blue Channel Spectrograph and 300 grating on 2017 June 29 UT. The data were reduced using standard procedures in \texttt{IRAF}; see Figure~\ref{fig:spec}.  We also show the original classification spectrum, available from the Transient Name Server \citep{xia2017b}.  For comparison, we plot the pre-maximum spectra of 5 well-studied SLSNe from the literature. The striking similarity in spectral slope and the characteristic \ion{O}{2} lines at $3500-4500$\,\AA\ supports the classification by \citet{dong2017}, and confirms that SN\,2017egm belongs to this class.

We arrange the spectra in Figure~\ref{fig:spec} in order of rest-frame time since explosion, using the best-fit explosion dates from the models of \citet{nic2017c}. The spectra were obtained up to 2 weeks before maximum light, but this corresponds to a range of times since explosion given the diversity in rise times; we discuss the estimated explosion date for SN\,2017egm in \S\ref{sec:fit}.  The first spectrum of SN\,2017egm is one of the earliest spectra ever obtained for a SLSN ($\approx 10$ d after explosion). After smoothing the data with a Savitsky-Golay filter, we recover the strongest \ion{O}{2} lines even at this early phase.

The centroids of the \ion{O}{2} absorption lines exhibit a clear trend towards redder wavelengths with increasing time since explosion. In the case of SN\,2017egm, the implied velocity decreases from $\approx 19,000$ \kms\ to $\approx 13,000$ \kms\ over a period of 24 d. This is comparable to the rate of decrease in \ion{Fe}{2} velocity shortly after maximum light \citep{LiuModjaz16}, but it is the first time this has been observed at such an early phase, using \ion{O}{2} lines, when few spectra are typically available. 
By the time of our MMT spectrum, obtained close to optical maximum light, the velocity of the \ion{O}{2} blends has decreased sufficiently that individual components can be resolved (see inset).
This spectrum likely also shows blends of \ion{Fe}{2} and \ion{Fe}{3} \citep{nic2013,nic2016b,LiuModjaz16}, and is more developed in the red, with overall similar features to SN\,2015bn in particular \citep{nic2016b}.

We will show in \S\ref{sec:fit} that the earliest spectrum of SN\,2017egm may have coincided with an initial bump in the light curve. This would be the first time that a spectrum has been obtained for a SLSN in this phase. The similarity of the early spectrum to those at maximum light is an important clue to the mechanism powering the bumps.

\begin{figure}
\includegraphics[width=\columnwidth]{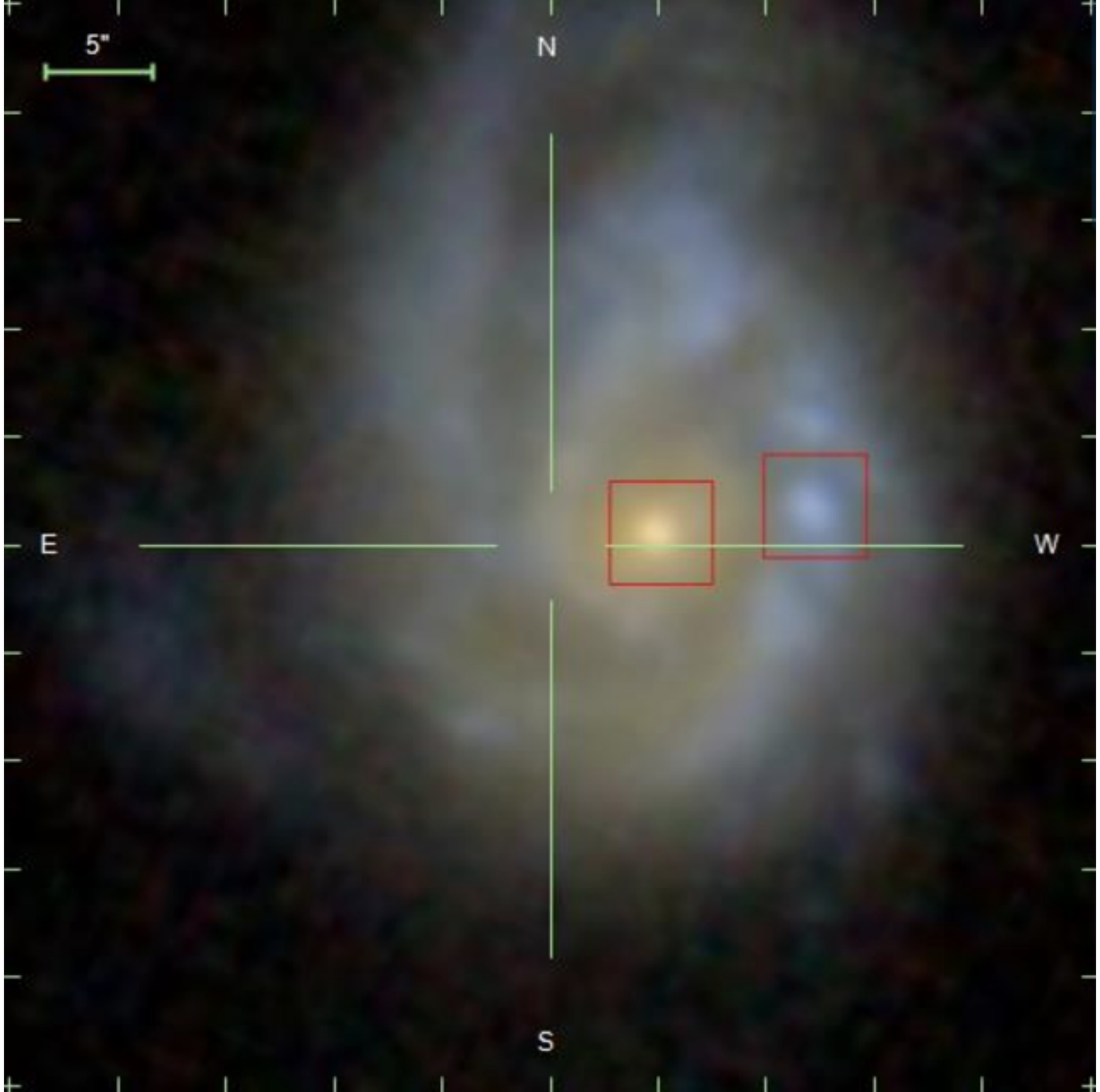}
\includegraphics[width=\columnwidth]{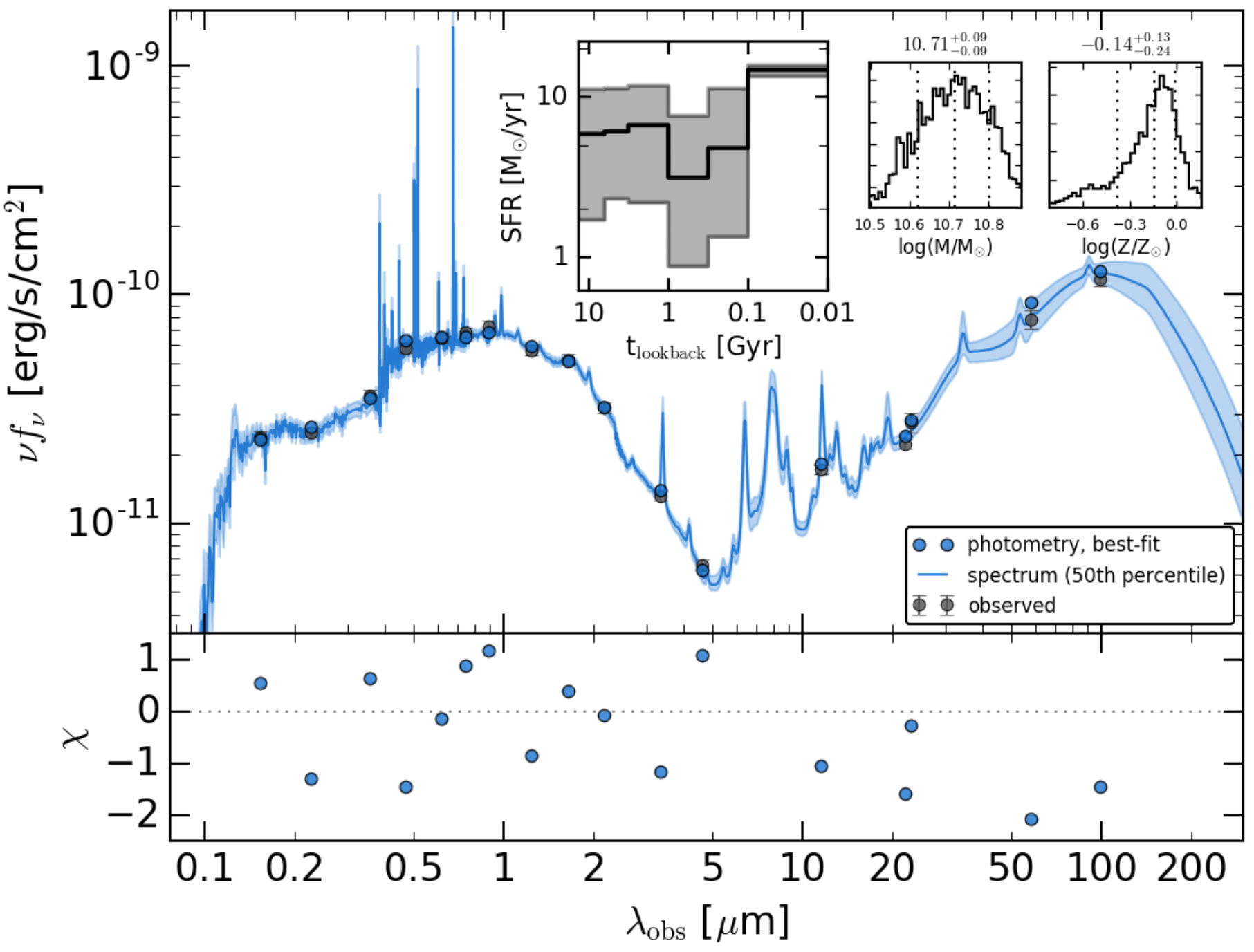}
\centering
\caption{{\it Top:} SDSS DR13 \citep{alb2016} image of NGC\,3191. The cross-hairs mark the position of SN\,2017egm. The boxes mark the positions of the SDSS fiber spectra from which we measured the metallicity; the red region is the galaxy nucleus, while the blue region to the west is a star-forming region at a similar radial offset from the nucleus as SN\,2017egm. {\it Bottom:} Fit to the spectral energy distribution of NGC\,3191 using \texttt{Prospector} \citep{leja2017}.  The inset plots show the star-formation history, as well as the stellar mass and metallicity posteriors.}
\label{fig:host}
\end{figure}

\section{Host Galaxy: NGC\,3191}
\label{sec:host}

Integrated photometry of NGC\,3191 is available from radio to UV. Using the near-UV flux density measured with the Galaxy Evolution Explorer \citep[GALEX;][]{mar2005}, we infer a star formation rate of ${\rm SFR}=1.4\times 10^{-28}L_{\rm \nu,UV}\approx 5$ M$_\odot$ yr$^{-1}$.  On the other hand, infrared measurements at 25, 70, and 100 $\mu$m from the Infrared Astronomy Satellite \citep[IRAS;][]{neu1984} yield ${\rm SFR}\approx 11$ M$_\odot$ yr$^{-1}$ (using the relation of \citealt{rie09}).  NGC\,3191 also exhibits resolved emission at 1.4 GHz in data from the Faint Images of the Radio Sky at Twenty centimeters survey \citep[FIRST;][]{bec1995}, pointing to a star formation origin, with an inferred value of ${\rm SFR}\approx 13$ M$_\odot$ yr$^{-1}$ (using the relation of \citealt{yc02}).

\citet{gru2017} recently detected X-ray emission from NGC\,3191 using the X-ray Telescope (XRT) on \textit{Swift}. We analyze the XRT data spanning 2017 June 2--20  (exposure time of $19.2$ ks) and find a source with a count-rate of $(1.6\pm 0.4)\times 10^{-3}$ s$^{-1}$ at position RA=\ra{10}{19}{04.44}, Dec=\dec{$+$46}{27}{18.7}. This is about $13''$ away from the SN position, making a direct association unlikely. We fit the extracted counts with a power law spectral model using a Galactic column of $N_{H,{\rm MW}}=9.6\times 10^{19}$ cm$^{-2}$ \citep{wil2013}. We find no evidence for intrinsic absorption ($N_{H,{\rm int}}\lesssim 0.7\times 10^{22}$ cm$^{-2}$; $3\sigma$). The power law index is $\Gamma=2.2\pm 0.4$, leading to an unabsorbed flux of $6.2^{+1.8}_{-1.2}\times 10^{-14}$ erg cm$^{-2}$ s$^{-1}$ ($0.3-10$ keV) or $L_X\approx 1.3\times 10^{41}$ erg s$^{-1}$.  Interpreting this luminosity as due to star formation yields a value of ${\rm SFR} \approx 10$ M$_\odot$ yr$^{-1}$ (converting to the $2-10$ keV band and using the relation of \citealt{ggs03}), in excellent agreement with the IR and radio inferred values.  We therefore conclude that NGC\,3191 has a total star formation rate of $\approx 15$\,M$_\odot$\,yr$^{-1}$, of which about $70\%$ is dust obscured, and that the X-ray emission detected with XRT is likely not dominated by SN\,2017egm itself. 

We further fit the broad-band spectral energy distribution of NGC\,3191 using {\tt prospector} \citep{leja2017} to determine additional galaxy parameters (Figure~\ref{fig:host}). The results indicate a star-forming galaxy with a moderate amount of dust and no evidence for AGN activity. The stellar mass is $\log M_* = 10.7 \pm 0.1$\,\M, and SFR\,$=14.8 \pm 1.2$\,\M\,yr$^{-1}$, in excellent agreement with our earlier estimates. The integrated metallicity is $0.7\pm 0.3$\,\Z\footnote{Note that the metallicity grid in \texttt{Propsector} extends only to 1.5\,\Z}.

The half-light radius of NGC\,3191 as determined from Sloan Digital Sky Survey \citep[SDSS;][]{alb2016} $u$-band imaging is $6.9\pm 0.2''$. The host-normalized offset of SN\,2017egm from the nucleus is therefore $R/R_{\rm half}\approx 0.75$. This is close to the median for the SLSN sample \citep[1.0;][]{lun2015} and the LGRB sample \citep[0.63;][]{bla2016}.

Finally, we use archival SDSS spectra taken both at the host nucleus and at the location of a bright star forming region about $7''$ away from the center to infer the gas-phase metallicity (Figure~\ref{fig:host}). The line ratios $\log ($[\ion{N}{2}]/H$\alpha) = -0.40$ and $\log ($[\ion{O}{3}]/H$\beta) = -0.35$ are typical of SDSS star-forming galaxies with minimal contribution from an AGN \citep{bal1981,kew2006}. For the purpose of comparison with previous metallicity measurements for SLSN hosts, we use the calibration of \citet{kob2004}, utilizing both the [\ion{N}{2}]/H$\alpha$ ratio and the $R_{23}$ diagnostic. We find that at the galaxy center the metallicity is $12+{\rm log(O/H)}\approx 9.0$ ($\approx 2$ Z$_\odot$), while at the location of the star forming region it is $12+{\rm log(O/H)}\approx 8.8$ ($\approx 1.3$ Z$_\odot$), indicative of a mild gradient.  While we do not have a metallicity measurement directly at the SN position, its radial offset from the center is comparable to that of the star forming region (see Figure \ref{fig:host}) and we therefore conclude that it is most likely $\gtrsim {\rm Z}_\odot$.

\begin{figure}
\includegraphics[width=\columnwidth]{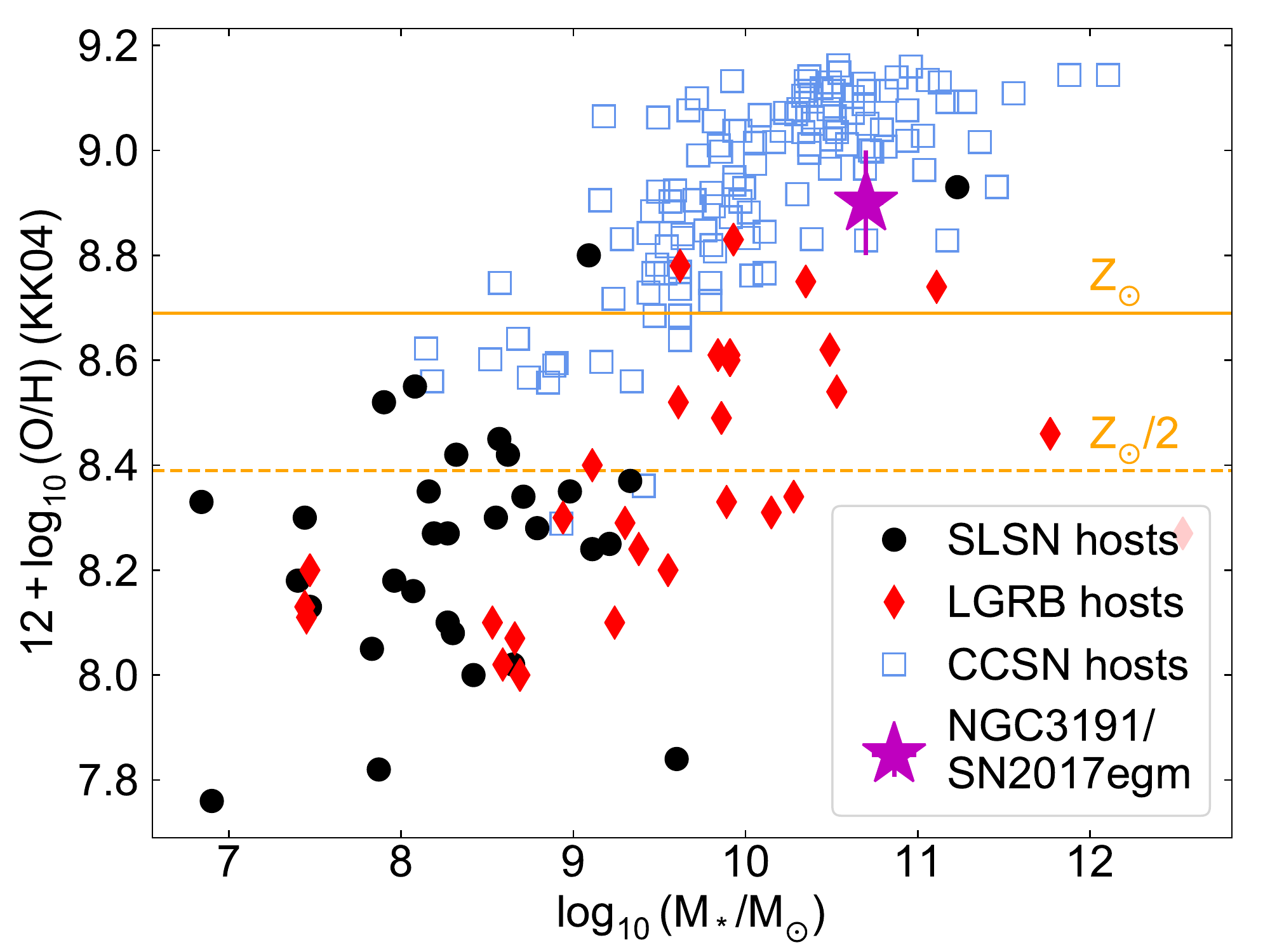}
\caption{Mass-metallicity diagram for SLSN (black circles, median $z=0.25$), LGRB (red diamonds, median $z=0.80$) and core-collapse SN (blue squares, median $z=0.66$) host galaxies.  The host of SN\,2017egm (NGC\,3191) is shown as a magenta star. The error bar in metallicity represents the range between the host nucleus and the star-forming region in SDSS (see text). About $10\%$ of SLSNe and LGRBs appear to occur at Solar metallicity.  The comparison data are from \citet{lun2014,lel2015,per2016,chen2016,sav2009,lev2010,sve2010,gra2013,per2016b,kru2015,schu2016}.}
\label{fig:Z}
\end{figure}

In Figure~\ref{fig:Z} we show NGC\,3191 on the mass-metallicity diagram, along with previous SLSN, LGRB, and core-collapse SN host galaxies.  Clearly, NGC\,3191 has a higher metallicity and stellar mass than most previous SLSN hosts, comparable to those of regular core-collapse SN hosts.  However, similar cases exist in the LGRB sample, and two other SLSNe appear to have metal-rich hosts: MLS121104 \citep{lun2014} and PTF10uhf \citep{per2016}. Both were at $z\approx 0.3$ and occurred in the outskirts of their host galaxies, and therefore it is not possible to exclude significant metallicity differences between the integrated host and the explosion site, or even the presence of undetected dwarf satellites as the true hosts (e.g., \citealt{chen2016b}). \citet{per2016} found that PTF10uhf was coincident with a merger between two star-forming galaxies, and may be more likely associated with the less massive of the pair. However, the metallicity they inferred from a spectrum extracted at the SN location did not indicate a significantly lower metallicity there. For SN\,2017egm, due to its low redshift and small offset from the host nucleus, a Solar metallicity is more robust.  This in turn lends support to the possibility that MLS121104 and PTF10uhf also occurred in Solar metallicity environments.

We therefore conclude that $\sim 10\%$ of SLSNe may occur in galaxies with $Z\gtrsim {\rm Z}_\odot$, while the bulk of the sample remains skewed to lower metallicities (e.g., \citealt{lun2014,per2016,chen2016,schu2016}); this is similar to the case for LGRBs \citep{lev2013,gra2013,per2016b}.  While the precise fraction of SLSNe observed in metal-rich galaxies will be sensitive to selection effects and detection efficiencies, our results indicate that there is no hard upper bound on the metallicity of SLSN environments.

\section{Light curve analysis}
\label{sec:fit}

Upon the classification of SN\,2017egm as a SLSN, \textit{Swift} UV-Optical Telescope (UVOT) imaging was obtained starting on UT 2017 June 2 \citep[P.I. Dong; see][]{dong2017}. We downloaded the data from the public \textit{Swift} archive and extract the UVOT light curves in the \textit{UVW2}, \textit{UVM2}, \textit{UVW1}, \textit{U}, \textit{B}, and \textit{V} filters following the procedures outlined in \citet{bro2009}, using a $3''$ aperture to minimise contamination from the underlying host galaxy light.  The magnitudes are calibrated in the \textit{Swift} photometric system \citep[Vega mags;][]{bre2011}. We estimate the host contamination by extracting the flux in a $3''$ aperture centered on a bright part of the galaxy far from the SN position, and subtract this contribution from the UVOT photometry. Given the resulting surface brightness of $m_u\approx 23$\,AB\,mag\,arsec$^{-2}$ the host contribution is only a few percent and it therefore has a minimal effect on the light curves and our models. We also obtained $g$, $r$, $i$ imaging with the FLWO 48-inch telescope and the 1.3-m McGraw-Hill telescope at MDM observatory.

\begin{figure*}
\centering
\includegraphics[width=8cm]{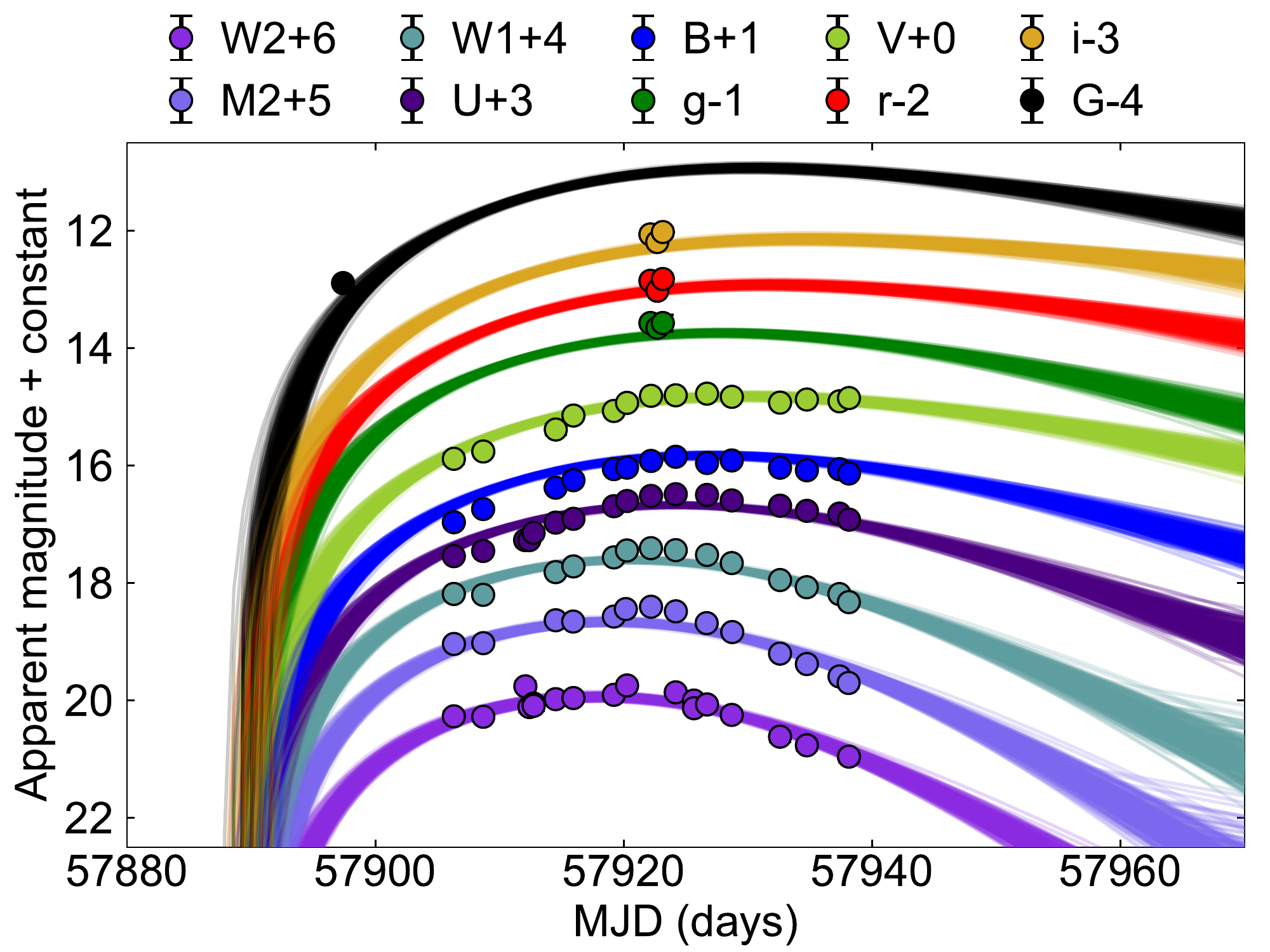}
\includegraphics[width=7cm]{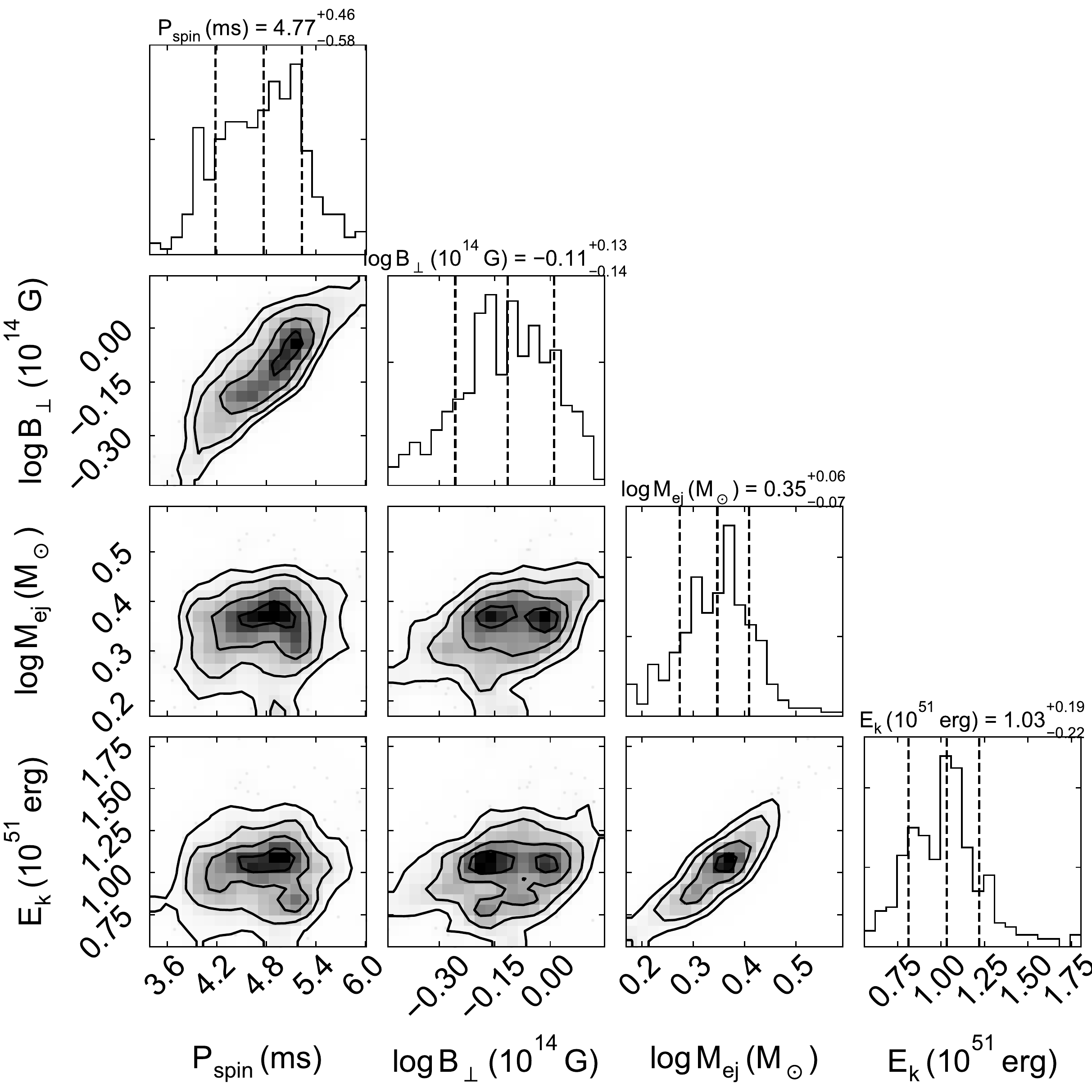}
\includegraphics[width=8cm]{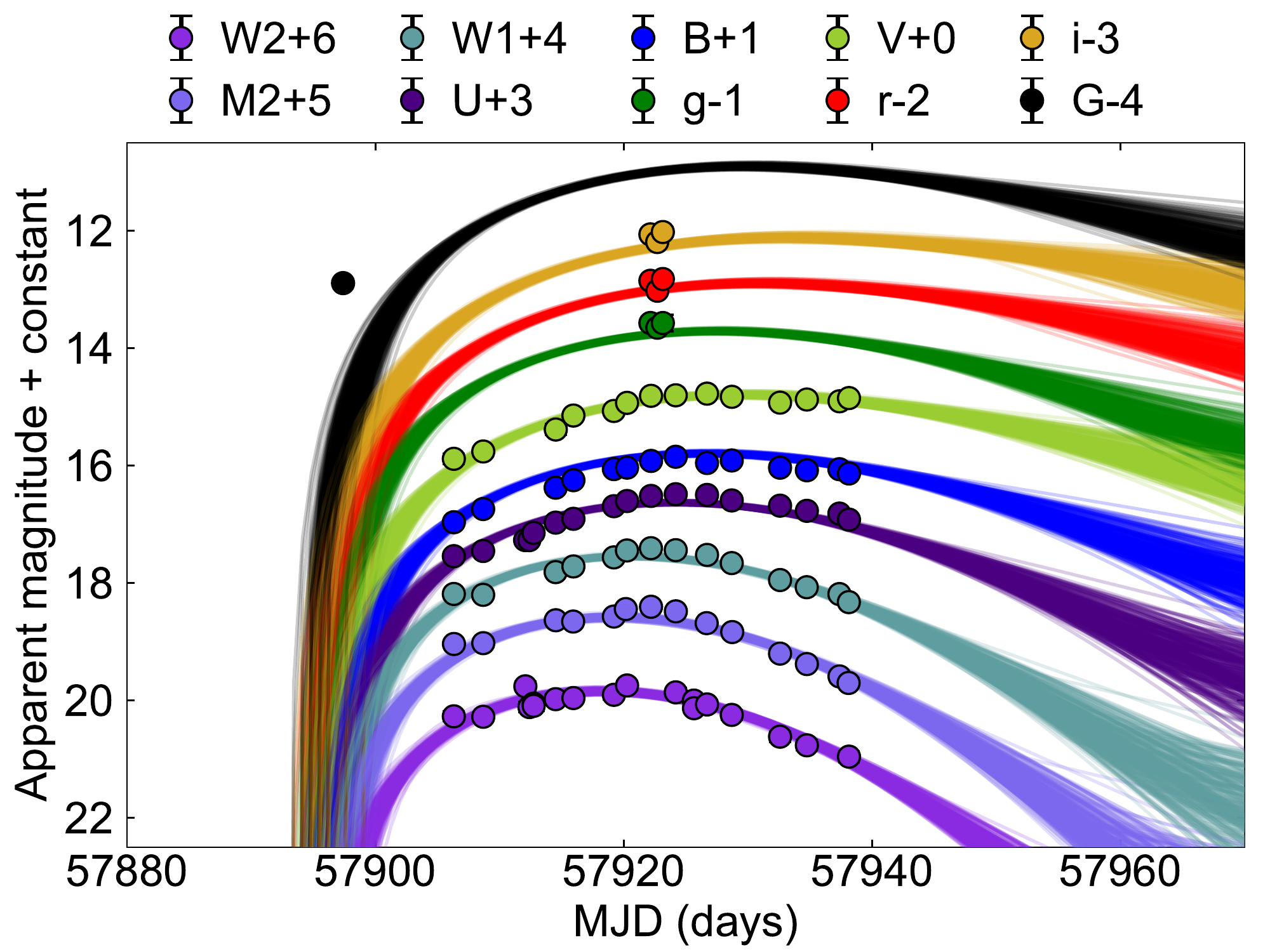}
\includegraphics[width=7cm]{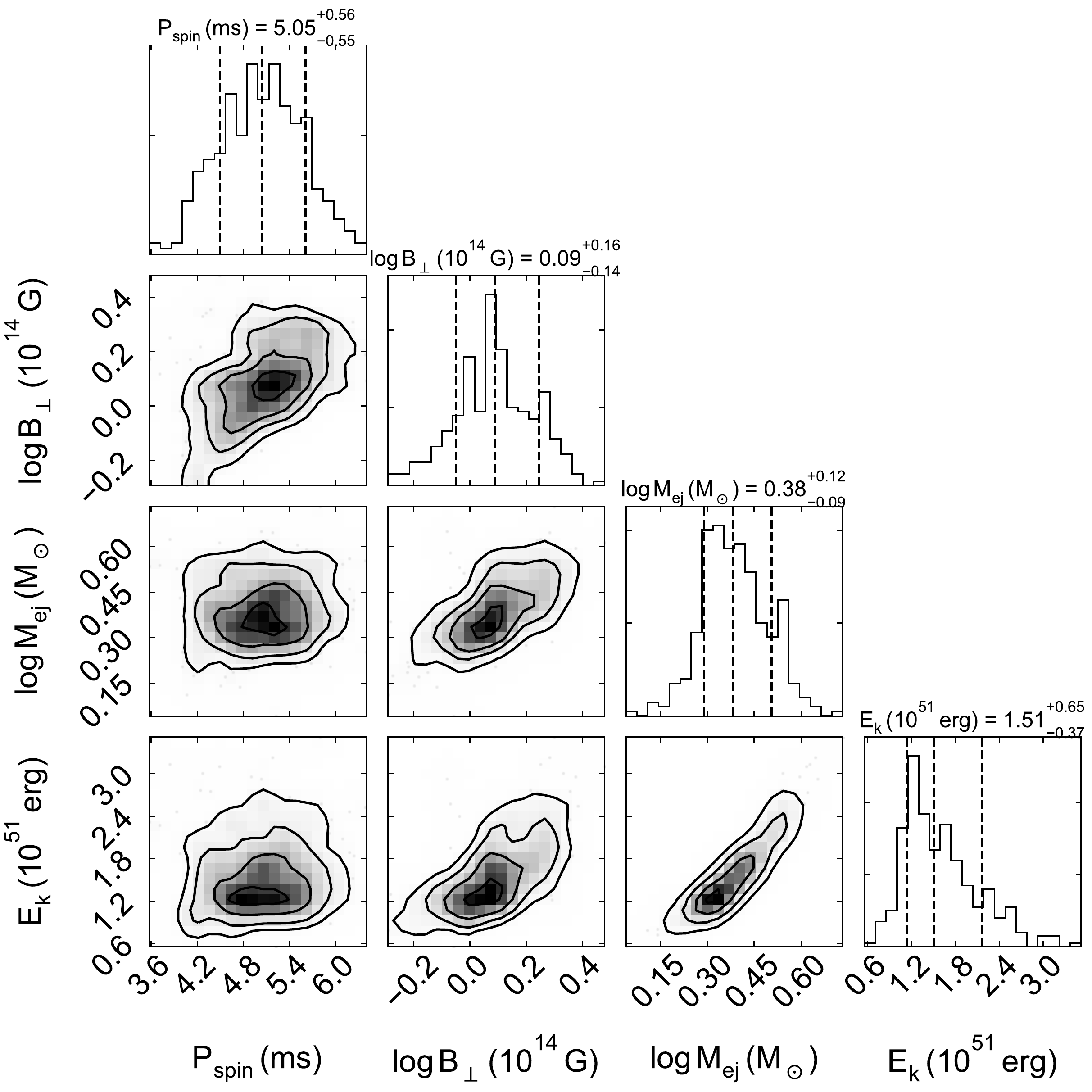}
\includegraphics[width=8cm]{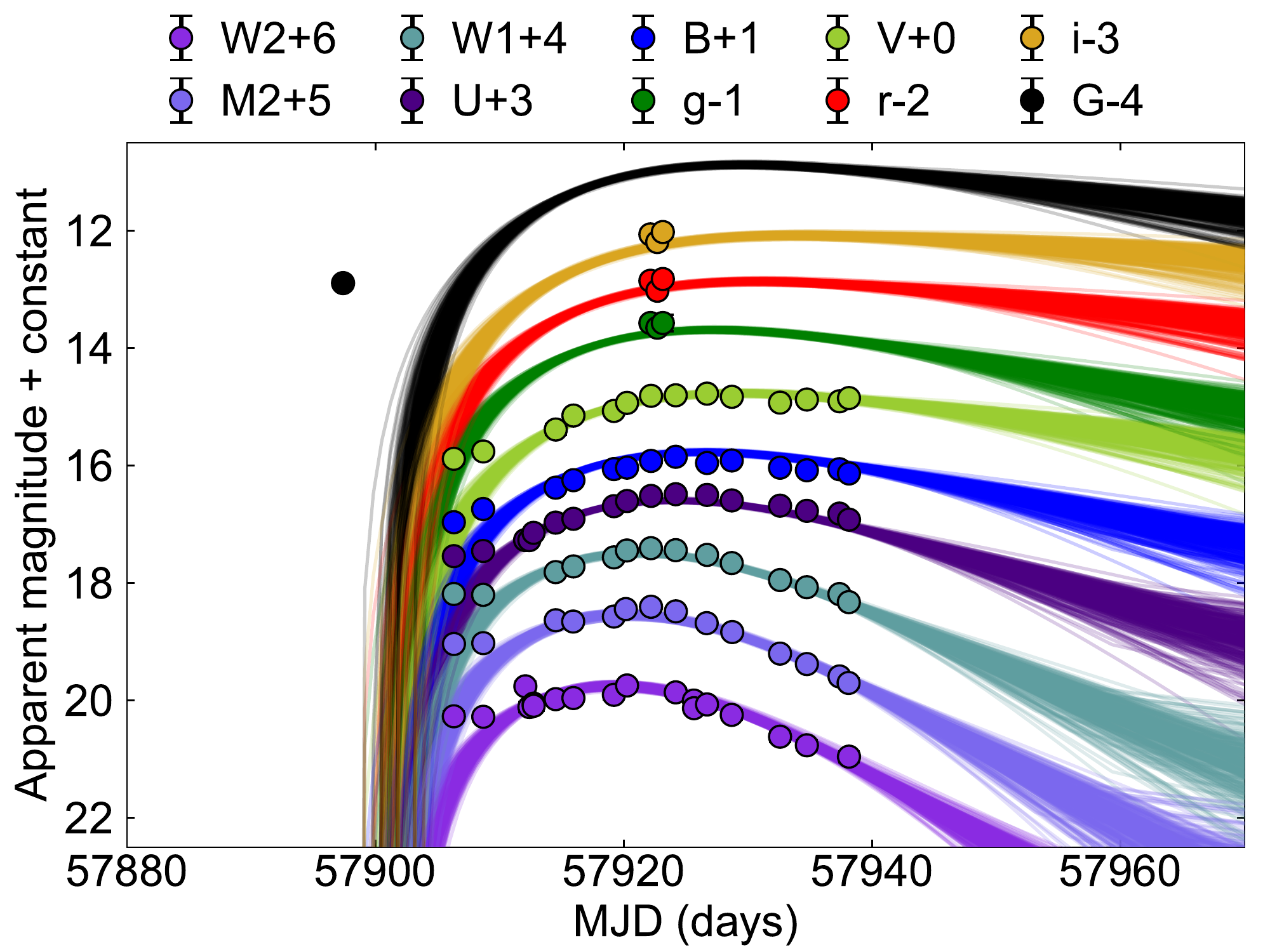}
\includegraphics[width=7cm]{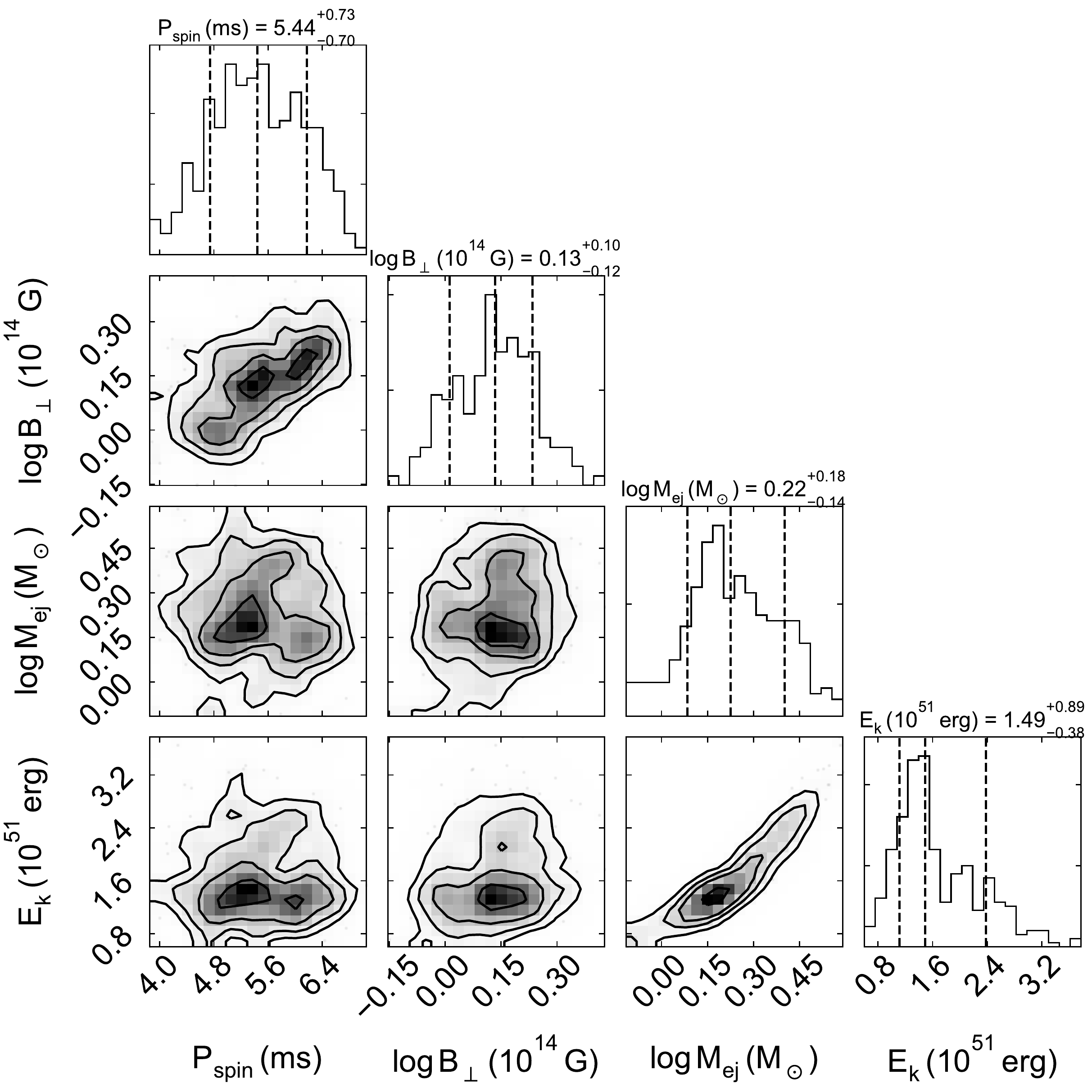}
\caption{Ensembles of magnetar model light curve fits to SN\,2017egm (\textit{left}) and derived posteriors (\textit{right}). \textit{Gaia} data are in the $G$ filter (central wavelength 6631\,\AA). \textit{Top}: Fits including all of the available data (Model 1). \textit{Middle}: Fits excluding the \textit{Gaia} data point (Model 2). \textit{Bottom}: Fits excluding the \textit{Gaia} data point and the two earliest epochs of UVOT data (Model 3).}
\label{fig:lc}
\end{figure*}

We fit the multicolour UVOT light curves and the \textit{Gaia} data point using the Modular Open Source Fitter for Transients (\mosfit). Specifically, we use the magnetar-powered model recently fit to a sample of 38 SLSNe by \citet{nic2017c}, with identical distributions of priors.  The model and implementation in \mosfit, and results for the SLSN population, are described in detail by \citet{nic2017c}; a fuller description of the code and its usage will be provided by Guillochon et al.~(in preparation).

The earliest light curve data for SLSNe are often complicated by the presence of fast initial peaks, or `bumps' around the time of explosion \citep{lel2012,nic2015a,nic2016a,smi2016}. These have been variously interpreted as shock-cooling of extended material \citep{piro2015}, a secondary shock driven by the engine \citep{kas2016}, or most recently a shocked `cocoon' surrounding a jet breakout \citep{mar2017}. However, our magnetar model for the main light curve peak does not accommodate these bumps explicitly. Hence we test for a bump in SN\,2017egm by fitting the models to three versions of the light curves: with all data included (i.e., no bump; Model 1); with the \textit{Gaia} data point excluded assuming that it represents a rapid bump phase (Model 2); and with the \textit{Gaia} data point and first two UVOT epochs excluded assuming that they represent an extended bump (Model 3). All of these models and their associated posteriors are shown in Figure \ref{fig:lc}.

\subsection{Observational properties}

SN\,2017egm reached maximum light in $U$-band on MJD 57924 with $m_U \approx 13.5$\,mag, or $M_U\approx -22.2$\,mag. The apparent magnitude is $\gtrsim 2$\,mag brighter than any previous SLSN due to the proximity of SN\,2017egm, but the absolute magnitude is typical for SLSNe.

We use our light curve models to estimate the explosion date and rise time to maximum light. The explosion date varies between our models depending on the assumption of a pre-explosion bump. If we assume the rise is smooth (Model 1), we find an explosion date of MJD $57889\pm 1$ (2017 May 16).  In Model 2 we find MJD $57895\pm 1$ (2017 May 22). In Model 3, we find MJD $57901\pm 1$ (2017 May 28), which is after the \textit{Gaia} detection -- in this case a precise estimate of the explosion date would require a detailed model for the bump itself (since the rise in our model is always monotonic).  Overall, the range of rise times corresponding to the various model fits is $\approx 27-34$ days, consistent with typical values for SLSNe \citep[e.g][]{nic2015b,nic2017c}.

Comparing the different models in Figure \ref{fig:lc}, we find that models that exclude points potentially contaminated by a bump (Models 2 and 3) provide somewhat better fits to the rise and maximum light behaviour. This is quantified in terms of a variance parameter \citep[the additional statistical uncertainty to be added uniformly to all data points such that the model fit has $\chi^2=1$; see][Guillochon et al.~in preparation]{nic2017c}. We find median values of $\sigma = 0.14$, 0.11 and 0.08\,mag for Models 1, 2 and 3, respectively. In the case of Model 3, this is similar to the observed photometric errors. Therefore there is modest evidence favouring the fits in which we assume a bump. Future data during the light curve decline will place further constraints on the model fits, which may help to emphasise the differences with and without a bump.

In the latter models, the light curve fits are well below the earliest points from \textit{Gaia} and UVOT, and the first points in the UV bands appear to show little evolution or even a slight dimming. This is fully consistent with the properties of the early bumps in SLSNe, and suggests that the \textit{Gaia} discovery may  have occurred during this phase.  We therefore encourage all surveys to search their archives for more data around this time.  

It is therefore possible that the spectrum from \citet{xia2017}, obtained only 3 days after discovery (Figure \ref{fig:spec}), is the first spectrum of a SLSN ever taken during the bump phase. It is therefore noteworthy that it is remarkably similar to the spectrum at maximum light, with high velocity lines and a smooth velocity evolution.  This  can inform models for the underlying mechanism of these bumps.

SN\,2017egm will soon enter Solar conjunction and will be unobservable between 2017 July 4 and September 16.  While this prevents monitoring over several months of post-maximum evolution, SN\,2017egm will remain bright enough for detailed study long after it becomes visible again. The late-time decline rate is not well constrained by our data, but our models provide probabilistic estimates. Assuming a limit of $\lesssim 23$\,mag for spectroscopy, our ensemble of fits suggests that SN\,2017egm will remain brighter than this limit for $\gtrsim 2$ years, enabling unprecedented deep observations at phases that have not been possible to probe for more distant SLSNe.

\subsection{Physical parameters: is SN\,2017egm unique?}

Given the  metal-rich environment of SN\,2017egm, it is imperative to explore whether this nearby event is representative of the general SLSN population, or if it differs in some fundamental properties. We can address this question with our light curve fit and the broad comparison sample of \citet{nic2017c}.

\begin{figure*}
\centering
\includegraphics[width=15cm]{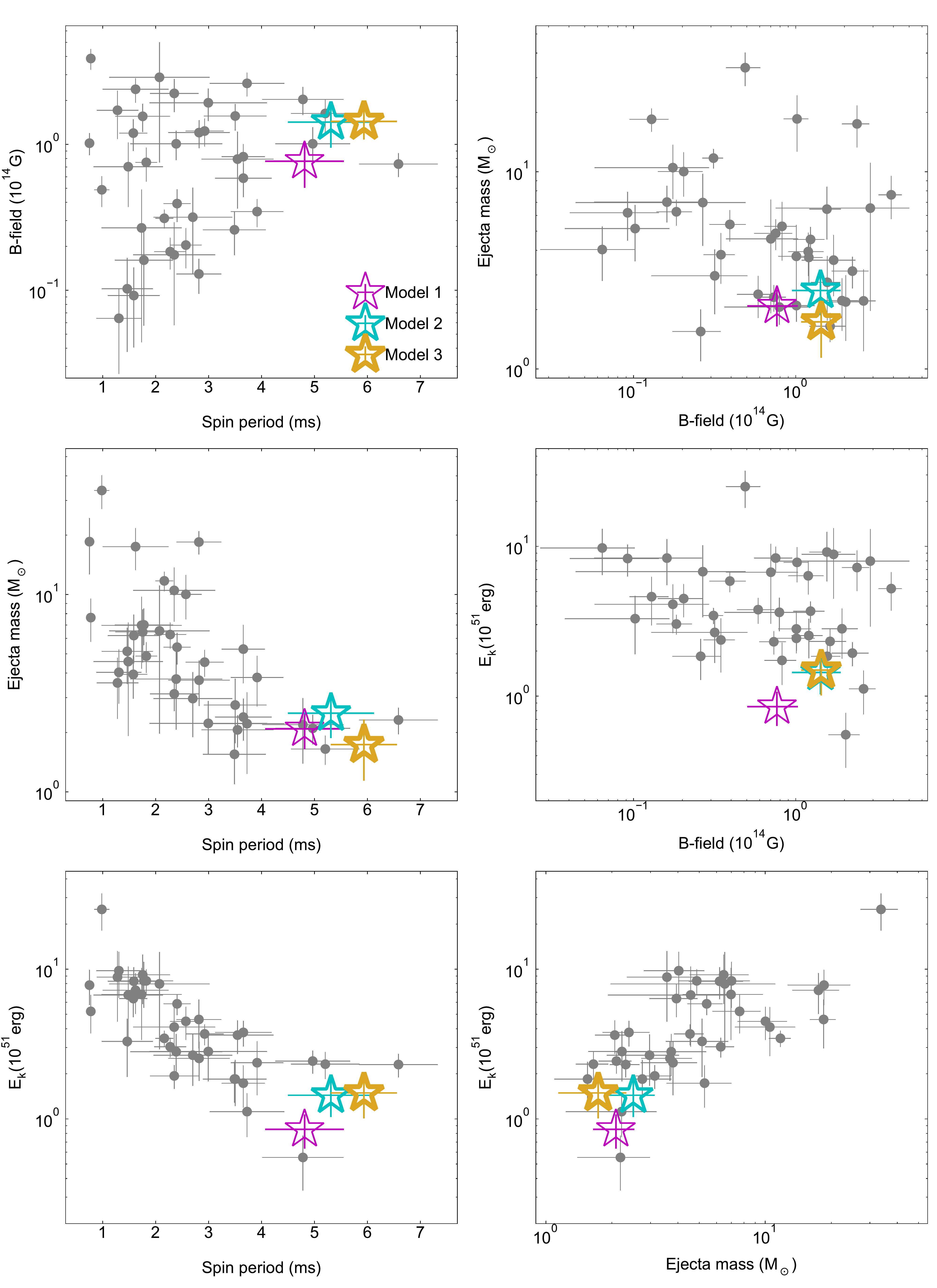}
\caption{Physical parameters of our magnetar fits compared to the sample of \citet{nic2017c}.  We show results for the three models shown in Figure~\ref{fig:lc}.  Regardless of the exact model SN\,2017egm overlaps with the full sample, demonstrating that this event does not differ significantly from those in low-metallicity environments.}
\label{fig:params}
\end{figure*}

In Figure~\ref{fig:params} we plot the medians and error bars for the ejecta mass ($M_{\rm ej}$), kinetic energy ($E_K$), magnetar spin period ($P$), and magnetic field ($B$) estimated from our three model fits to the existing data for SN\,2017egm, in comparison to the full SLSN sample (see Figure 5 of \citealt{nic2017c}).  We find $P\approx 4-6$\,ms, $B\approx (0.7-1.7)\times 10^{14}$\,G, $M_{\rm ej}\approx 2-4$\,\M, and $E_K\approx 1-2\times 10^{51}$\,erg.  These values are well within the distribution for the overall SLSN sample (all are within $\approx 1\sigma$ of the population medians), indicating that SN\,2017egm is  not an atypical event. Thus, despite its occurrence in a metal-rich environment, SN\,2017egm appears to be a typical member of the SLSN population.

We do however note that the relatively low ejecta mass and modest spin period we infer could in principle be a reflection of the environment, in the sense that higher metallicity may lead to greater loss of mass and angular momentum before explosion.  \citet{chen2016} found a possible correlation between host metallicity and magnetar spin period, although \citet{nic2017c} did not find such a trend with a larger sample of events.  A larger number of SLSNe at the high-metallicity end will be needed to test if there does exist a dependence of spin period or ejecta mass on metallicity.  Given that it overlaps the rest of the population at the $\approx 1\sigma$ level in all parameters, SN\,2017egm does demonstrate that any such dependence is weak. Thus, the main effect of metallicity is a reduction in the rate of occurrence of SLSNe, rather than a change in the explosion properties.

\section{Conclusions}

We have analysed the host galaxy and early evolution of SN\,2017egm --- the nearest SLSN to date and the most securely associated with a Solar-metallicity environment.  Our analysis makes predictions for the explosion time and future light curve evolution of SN\,2017egm, suggests that the early data may represent a `bump' phase, and demonstrates that despite the higher metallicity the properties of SN\,2017egm are typical of the overall SLSN sample.

Using archival photometry and spectroscopy of NGC\,3191 we find $M_*\approx 10^{10.7}$ M$_\odot$, ${\rm SFR}\approx 15$\,\M\,yr$^{-1}$, and a metallicity at a comparable radial offset to that of SN\,2017egm of $\approx 1.3$\,\Z ($\approx 2$\,\Z at the nucleus). Together with two other SLSNe from the literature that appear to be in similar galaxies (but at $z\approx 0.3$), we estimate that up to $\approx 10\%$ of SLSNe may occur at Solar metallicity, similar to the findings for LGRBs.  While the SLSN rate is clearly suppressed at high metallicity \citep{per2016,chen2016,schu2016}, there does not seem to be a strong upper bound on the metallicity.

We model the pre-maximum UV and optical photometry with \mosfit, assuming a magnetar central engine, and find parameters typical of the general SLSN population \citep{nic2017c}. We use the ensemble of fits to make probabilistic estimates for observables such as the explosion date and the time for which SN\,2017egm will be observable; SN\,2017egm should be observable spectroscopically for $\gtrsim 2$ years, allowing for future detailed studies of unprecedented detail.

If we assume a monotonic rise, we estimate an explosion date of MJD $57889  \pm 1$\,d. This will inform archival searches for any pre-maximum `bumps' around the time of explosion \citep{nic2016a}.  In fact, our modeling already shows some discrepancies with the data at the earliest epochs that likely indicates a bump has occurred. If confirmed, this makes the spectrum obtained by \citet{xia2017} the first ever obtained during the bump phase of a SLSN. We found that that this spectrum matches typical SLSN spectra at maximum light, though with \ion{O}{2} lines that are more strongly blue-shifted.

The most important conclusion of our study is that metallicity has at most a modest effect on the physical parameters of SLSNe and their engines, and primarily impacts on the overall rate. This supports the analysis of \citet{nic2017c}, who found no correlations between metallicity and any model parameters for their SLSN sample. A similar conclusion has been reached for LGRBs \citep{lev2010b}. However, it is also possible that the relatively slow spin period (while still overlapping the rest of the spin distribution for SLSNe) is a reflection of the metal-rich environment \citep{chen2016}.

Continued multi-wavelength observations of SN\,2017egm will constrain the parameters more tightly.  We will continue to model the data and distribute results via the Open Supernova Catalog \citep{gui2017}, providing a test for whether appropriately calibrated physical models with \mosfit can provide robust predictions to aid in follow-up observations of SN\,2017egm and future SLSNe and other transients. If SN\,2017egm continues to evolve in a similar fashion to other SLSNe, it will support a picture where fairly normal SLSNe can sometimes occur in unexpectedly metal-rich environments. Future data will also tell us whether the closest SLSN to date has any more surprises in store.

\acknowledgments
We thank Iair Arcavi for important comments that improved this work. The Berger Time-Domain Group at Harvard is supported in part by NSF grant AST-1411763 and NASA ADA grant NNX15AE50G. This paper uses data products produced by the OIR Telescope Data Center, supported by the Smithsonian Astrophysical Observatory.
This work is based in part on observations obtained at the MDM Observatory, operated by Dartmouth College, Columbia University, Ohio State University, Ohio University, and the University of Michigan.

\end{document}